\documentclass[twocolumn,preprintnumbers,nofootinbib,prl]{revtex4}

\usepackage{graphicx}

\usepackage[hypertex]{hyperref}
\newcommand{\beq}{\begin{equation}}
\newcommand{\eeq}{\end{equation}}
\newcommand{\be}{B_\oplus}

\def\be{\begin{equation}}
\def\ee{\end{equation}}
\def\baray{\begin{eqnarray}}
\def\earay{\end{eqnarray}}
\def\ba{\begin{eqnarray}}
\def\ea{\end{eqnarray}}

\begin{document}

\pagestyle{plain}

\preprint{MAD-TH-06-8}

\title{Observing the Geometry of Warped Compactification via Cosmic Inflation}

\author{Gary Shiu and Bret Underwood}

\affiliation{Department of Physics, University of Wisconsin,
Madison, WI 53706, USA}


\begin{abstract}

Using DBI inflation as an example, we demonstrate that the detailed geometry
of warped compactification can leave an imprint on the cosmic microwave background (CMB).  We compute CMB
observables 
for DBI
inflation in a generic class of warped throats and find that the results
(such as the sign of the tilt of the scalar perturbations and its running) 
depend sensitively  on the precise 
shape
of the warp factor.  In particular, we 
analyze
 the warped deformed conifold
and find that the results can differ from those of other warped geometries,
even when these geometries approximate well the exact metric of the warped deformed conifold.

\end{abstract}
\maketitle

It has long been recognized that observational
cosmology provides a powerful window for testing fundamental physics, 
especially theories
whose energy scales are far from what current accelerator experiments can reach.
In this regard, inflation is a particularly gratifying idea. In addition to the strong and growing experimental
evidence in support of its basic picture, inflation significantly highlights the importance of microphysics.
Quantum fluctuations in the early universe are stretched by the enormous expansion of inflation to scales of astrophysical relevance, providing the seed for 
density perturbations.
Thus, by way of inflation, 
short distance physics may leave an imprint on cosmological observables such as the CMB \cite{Greene:2005aj}.

In recent years, significant progress has been made in constructing inflationary models from string theory. A particularly promising framework is the idea of
brane inflation \cite{DvaliTye,DDbar}, in which the attraction between the branes gives rise to the inflaton potential.
Advances in stabilizing moduli with fluxes \cite{ModuliStabilization,KKLT} have enabled
the realization of this idea; some explicit models can be found in \cite{KKLMMT,DBI}.
It is therefore worthwhile to explore if any stringy features, such as the geometry of compactification,
can be determined or distinguished from precision cosmology.

In this paper, we focus on the relativistic limit of brane inflation, known as DBI inflation \cite{DBI,DBITip,ShanderaTye}, because of its strong observational potential. In particular, the primordial non-gaussianity of the CMB in DBI inflation can be much
larger than in slow roll models and is within the reach of current and
next-generation CMB experiments like WMAP and Planck \cite{DBI,NonGauss}.  More importantly, the shape of non-gaussianities
is distinct \cite{NonGauss}.
Thus, non-gaussianities of such size and shape, if observed, can put interesting bounds on the mass of the inflation which from \cite{Baumann:2006th} (see also \cite{Berg:2004ek}) may in turn
constrain some gross features of the compact geometry.

A crucial element in the construction of brane inflation models are explicit solutions of
warped throats, the most well studied example being the warped deformed conifold (or the Klebanov-Strassler (KS) throat) \cite{KS}. 
Besides sustaining inflation by flattening the inflaton potential 
\cite{KKLMMT} or limiting the speed of the branes \cite{DBI},
significant warping is essential in reheating the universe \cite{reheating}.
Far from the tip of the throat and the bulk of the compact space (usually taken to be a Calabi-Yau manifold), the warped deformed conifold can be approximated by a product of $AdS_5 \times T^{1,1}$ (up to a logarithmic correction \cite{KT}),
where $T^{1,1}$ is a 5-dimensional Einstein-Sasaki manifold with the topology of $S^2 \times S^3$.
Therefore, as in \cite{RS}, the warp factor can 
generate a hierarchy of scales.
However, unlike an exact AdS throat, the warp factor approaches a constant  near the tip which can be understood from AdS/CFT correspondence as 
a confining phase in the infrared of the dual 
gauge theory.
In fact, the gauge/gravity duality suggests 
the existence of various other warped throats which are asymptotically AdS
but differ in their geometries near the tip, though explicit metrics for such other warped geometries
are not yet known at present.
A question of significance is: could 
the precise geometry of warped throats 
be distinguished by observations?

The purpose of this Letter is to answer this question affirmatively.
Naively, if the observed density perturbations 
leave the horizon in the AdS region of a warped throat (which, with the exception of
\cite{DBITip}, is assumed), 
small corrections to the geometry due e.g. to 
the details of the tip region are irrelevant.
However, we will show, contrary to this expectation, that even if the last 55 e-foldings of inflation takes
place in the AdS region, small differences in the geometry 
can leave an observable effect.

Consider a 10 dimensional geometry with a warped throat whose metric takes the form,
\begin{equation}
ds^2_{10} = f^{-1/2}(r)\eta_{\mu\nu}dx^{\mu}dx^{\nu} + f^{1/2}(r) (dr^2+ ds^2_{X_5})
\end{equation}
where $r$ is the proper distance from the tip of the throat and $f(r)$ is the warp factor.  
The angular
coordinates of $ds^2_{X_5}$ will not affect our analysis and we will ignore them from now on.  

The motion of a $D3$-brane in a warped space is given by the DBI action, which
in terms of the canonical scalar field $\phi = T_{D3}^{1/2} r$ and a rescaled
warp factor $\tilde{f}(r) = T_{D3}^{-1} f(r)$ is, 
\begin{equation}
S_{DBI} = -\int d^4x \left(\tilde{f}^{-1}\left(\sqrt{1-\tilde{f}\dot{\phi}^2}-1\right) - V(\phi)\right).
\label{eq:DBI}
\end{equation}
The factor inside the square root of Eq.(\ref{eq:DBI}) must always be positive, which leads to the speed limit
$\dot{\phi}^2 < \tilde{f}(\phi)^{-1}$ and a corresponding relativistic gamma factor 
$\gamma \equiv \frac{1}{\sqrt{1-\tilde{f}\dot{\phi}^2}} = \sqrt{1+4 M_p^4H'^2\tilde{f}(\phi)}$.  
DBI inflation occurs when the potential is steep, e.g., when the potential is dominated by a mass term
$V(\phi) = m^2\phi^2$, which
can arise for instance from moduli stabilization effects \cite{Baumann:2006th}.
In this case the Hubble parameter $H= m\phi(1-B\phi^2)/M_p$ where 
$B\phi^2$ represent 
small corrections coming from the kinetic term of the energy density \cite{ShanderaTye}, and
we have chosen to absorb a factor of $\sqrt{3}$ into the definition of $m$.
Notice that this implies $\gamma(\phi) \propto \tilde{f}^{1/2}(\phi)$ for relativistic ($\gamma \gg 1$) motion.

The number of e-folds can be written as
\begin{equation}
N_e = \int H dt = -\frac{1}{2 M_p^2}\int \frac{H}{H'} \gamma(\phi)\  d\phi
\label{eq:efoldFormal}
\end{equation}
where $'$ denote derivatives with respect to $\phi$.
Irrespective of $\gamma$, this leads to the standard 
Lyth bound \cite{Lyth:1996im} $\frac{d\phi}{dN_e} = M_p\sqrt{\frac{r}{8}}$ 
where $r = 16\epsilon_D/\gamma$ 
is the tensor to scalar ratio.

The scalar spectral index and its running can be written as \cite{ShanderaTye}
\begin{eqnarray}
n_s -1 &=& \frac{2 M_p^2}{\gamma}\left[-4\left(\frac{H'}{H}\right)^2+2\frac{H''}{H}+2\frac{H'}{H}\left|\frac{\gamma'}{\gamma}\right|\right]  \nonumber
\\
\frac{dn_s}{d\ln k} &=& \frac{d}{dN_e}n_s = \frac{2M_p^2}{\gamma}\frac{H'}{H}\frac{d}{d\phi}n_s
\label{eq:spect}
\end{eqnarray}
The first term of the spectral index is always negative and tends to make it red, while the last term is always positive and will 
tend to make it blue.  The middle term is proportional to the small corrections of the kinetic term.

Given the metric of a warped throat, Eq.~(\ref{eq:efoldFormal}),(\ref{eq:spect}) can in general be evaluated only numerically. We will present our numerical results for the KS throat, but to get an analytic idea that details of 
warped geometry can affect CMB observables, let us first consider a general warp factor of the 
form \cite{DBITip}:
\begin{equation}
\tilde{f}(\phi) = \frac{1}{f_0+f_2\phi^2 + f_4 \phi^4}
\label{eq:fgeneric}
\end{equation}
For such warp factor, the scalar spectral index Eq.(\ref{eq:spect}) can be written (without
any assumption for $\gamma=\sqrt{1+4M_p^4 H'^2\tilde{f}}$ and $H=m\phi/M_p$) ,
{\small
\begin{eqnarray}
n_s-1 = &&\frac{M_p}{m}\sqrt{\frac{f_0+f_2\phi^2+f_4\phi^4}{(1+\frac{f_0+f_2\phi^2+f_4\phi^4}{4M_p^2m^2})}}[-\frac{4}{\phi^2}+\nonumber \\
	&&\frac{2(f_2+2 f_4\phi^2)}{f_0+f_2\phi^2+f_4\phi^4} \left(\frac{1}{1+\frac{f_0+f_2\phi^2+f_4\phi^4}{4M_p^2 m^2}}\right)]
\label{eq:spectAnsatz}
\end{eqnarray}}
The differences between throats with different warped geometries can be seen readily from this equation.  In particular,
for a warped throat with a constant warp factor $\tilde{f}^{-1}\approx f_0$ the spectral index is red, while for
an AdS-like throat $\tilde{f}^{-1}\approx f_4\phi^4$ the spectral index is slightly blue (the contributions in
Eq.(\ref{eq:spectAnsatz}) cancel and the addition of
extra corrections to the energy density coming from the kinetic terms leads to a slightly blue spectrum \cite{ShanderaTye}).

Furthermore, the running of the scalar spectral index for $\gamma \gg 1$ (ignoring higher order terms in $H(\phi)$) can be written as,
{\footnotesize
\begin{equation}
\frac{dn_s}{d\ln k} = -\frac{2 M_p}{m}\frac{4 f_0^2+6
f_0f_2\phi^2+(f_2^2+8f_0f_4)\phi^4+2f_2f_4\phi^6}{\phi^4(f_0+f_2
\phi^2+f_4\phi^4)}.
\end{equation}}
The running is always negative except when $f_0=f_2=0$, in which
case higher order corrections in $H$ need to be included. These corrections
can lead to a small positive running in some cases.
We see,
then, that because the inflationary dynamics depend strongly on the warp factor through the speed-limiting behavior, details
of the warped geometry can show up in the inflationary observables.

\begin{figure}[t]
\begin{center}
\includegraphics[scale=1]{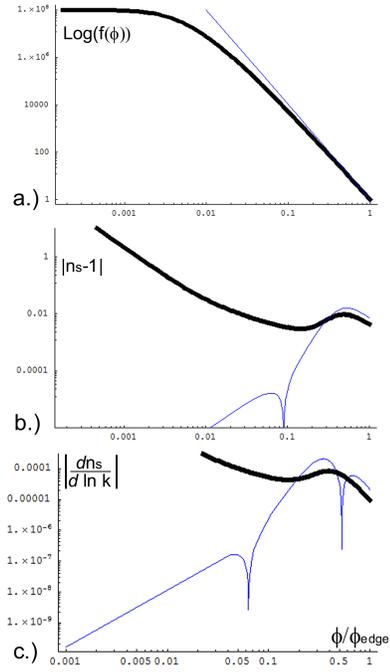}
\caption{\small a.) The warp factor for the AdS (blue dashed) and exact KS throat (black solid) are plotted as a function of the
canonical scalar field scaled by its value at the edge where the throat is glued to the bulk space.  We have chosen our parameters
to be $h_{tip}=10^{-2}$, $M_p = 100 m_s$ for these plots.
b.) The absolute value of the spectral index for the two throats (blue solid for AdS and thick black solid for KS).  
We take the absolute value of $n_s -1$ in order to show simultaneously
the results for AdS and KS throat in one plot.
The cusp in the AdS spectral curve corresponds to the tilt changing
from negative at larger $\phi$ to positive for smaller $\phi$.  This also happens where the warp factors
begin to differ.  
c.) The absolute value of the running of the scalar spectral index is shown for
the AdS (blue) and KS (thick black) throats.  The cusps in the AdS curve correspond to the running changing from negative to positive to 
negative again, as can easily be seen in the plot of the spectral index, see b.).}
\label{fig:KSAdSCompareTotal}
\end{center}
\end{figure}

\begin{figure}[t]
\begin{center}
\includegraphics[scale=1]{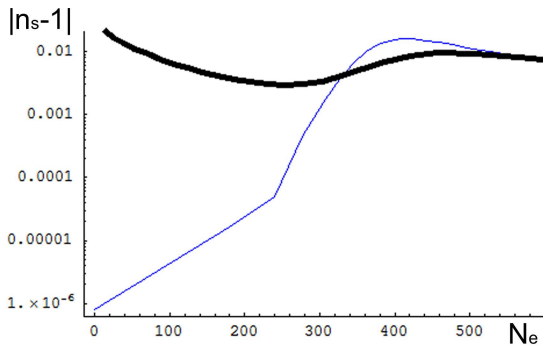}
\caption{\small The absolute value of the spectral index for the AdS (thin blue line) and KS (thick black line) throats is shown
as a function of the number of e-folds for $h_{tip} = 10^{-2}$ and $M_p = 100 m_s$.  The spectral index for AdS changes sign 
from negative to positive at the kinked point, so we
see that the region where spectral index for the two throats is different falls within the last 60 e-folds of inflation.}
\label{fig:KSAdSSpectNeCompare}
\end{center}
\end{figure}

To illustrate this effect more clearly, 
let us now turn to explicit warped throats in string theory.
For concreteness, we will 
consider two different warped throats and their corresponding warp
factors:
\begin{itemize}
\item AdS throat, with $\tilde{f} = \frac{\lambda}{\phi^4}$, cutoff at a fixed coordinate $\phi_{cutoff} = h_{tip} \lambda^{1/4} m_s$.
\item KS throat, with $\tilde{f} = 2^{2/3} (g_sM\alpha')^2\epsilon^{2/3}I(\tau(\phi))$, where $M$ corresponds
to the number of units of RR 3-form flux, $\tau(\phi)$ 
is a parameter along the radial direction of the throat, and $I(\tau(\phi))$ is an integral that can be determined numerically \cite{KS}.  
$\epsilon$ is related
to the deformation of the conifold and is not to be confused with the inflationary parameter $\epsilon_D$.
\end{itemize}
The KS geometry is sourced by a flux-induced D3 charge, so the KS throat asymptotically matches the 
AdS solution far from the tip \cite{KS}, see Figure \ref{fig:KSAdSCompareTotal}.  
For this reason, the AdS solution is often used as a simple model of the KS throat.
Near the tip of the throat the geometry is different, however, since the AdS
throat is cutoff at a finite value of the coordinate while the KS throat smoothly extends to $\phi=0$.
It is precisely this difference that will become pronounced in the inflationary observables.
The exact KS warp factor does not fit into our simple ansatz Eq.(\ref{eq:fgeneric}) and hence the corresponding inflationary observables have to be determined numerically. 
The ``mass gap" warp factor $\tilde{f} = \lambda (\mu^2+\phi^2)^{-2}$, where 
$\mu \propto \epsilon^{2/3}$ is related to the deformation of the
tip, is of this form and can be shown to be a reasonably good approximation to the KS throat \cite{DBITip}.
We can in principle compare the inflationary observables
for a AdS throat with that of the mass gap warp factor to simplify our calculations. 
However,
we wish to emphasize that the observable
effects of the throat geometry are present in 
some explict solutions of string theory
so
we will carry out our numerical 
analysis for the exact KS metric.

For the KS throat, the first term in the spectral index in Eq.(\ref{eq:spect}) dominates over the last, 
particularly near the tip where $\gamma'\propto f'\rightarrow 0$, so the spectral
index is red.  For the AdS throat the last term is slightly bigger than the first term due to the fact that 
$\gamma'$ keeps increasing
near the tip, so the spectral index is blue near the tip, see Figure \ref{fig:KSAdSCompareTotal}. Note that the differences
even become apparent 
when the deviation between the warp factors is small, so the details of the geometry can be important.
The running of the scalar spectral 
index can also be computed (the tensor spectral index and its tilt are not signficantly different),
and one can see that the magnitude and sign of the running strongly depend on the geometry of the throat.  We also show the spectral
index for the two different throat geometries as a function of the number of e-folds in Figure \ref{fig:KSAdSSpectNeCompare}, where one sees that the region
of the throat where 
the inflationary observables are different
includes the last 60 e-folds of inflation.

\begin{figure}[t]
\begin{center}
\includegraphics[scale=1]{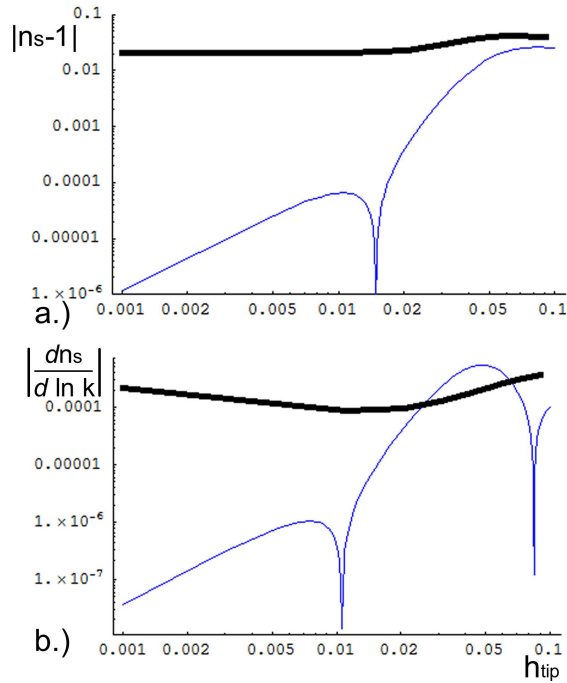}
\caption{\small a.) The absolute value of the scalar spectral index is shown for the AdS (thin blue line) and exact KS 
(thick black line) throats as a function of the
redshift factor at the tip $h_{tip}$ at 55 e-folds back from the end of inflation defined in the text.  We have fixed
$M_p = 100 m_s$.  b.) The running of the
spectral index also evaluated at 55 e-folds back for the AdS and KS throats.}
\label{fig:KSAdSSpect55Compare}
\end{center}
\end{figure}

In order to make comparisons between these throats, one must compare the values of the spectral indices and their running
when the observable modes cross the horizon.  Inflation should end for the KS throat when stringy effects become important
$\phi\sim \phi_s = h_{tip}m_s$ and for the AdS throat when the inflaton reaches the cutoff $\phi_{cutoff} = h_{tip} \lambda^{1/4} m_s$.
We have (numerically) evaluated the inflationary observables 55 e-folds back from these cutoffs for each throat, 
and the results are displayed in Figure \ref{fig:KSAdSSpect55Compare}.  It is then clear that the different throat geometries 
can have an observable effect
on the spectral index and its running.

Incidentally, we have also found that the spectral index for DBI inflation in KS throat is red tilted
(inclusion of the logarithmic correction term to the AdS throat as in \cite{KT} also leads to a red tilt).
Although we have only shown our results for $n_s -1$  as a function of $h_{tip}$ in 
Figure \ref{fig:KSAdSSpect55Compare}, the redness of the spectrum is robust
against the change of the other parameter $m_s/M_P$.
It is clear from Figure \ref{fig:KSAdSCompareTotal} that the spectral index
can be blue only for sufficiently small $\phi$. In this region of the throat, a scaling of $m_s/M_P$ is equivalent to a scaling of $h_{tip}$.

Finally, let us briefly comment on the observability of tensor modes.  For the KS throat, we find that the tensor to scalar ratio $r$ is too small to be observed unless the inflaton moves Planckian distance as suggested by the Lyth bound.  In contrast, the tensor signal is naively
observable in an AdS throat for some ranges of parameters \cite{ShanderaTye}, although
whether models with observable $r$ are under control within effective field theory is a subject of active investigation \cite{LythST}. 
Nonetheless, it is clear that 
one should be cautious in using the 
AdS throat as an approximate description for 
more realistic warped geometries.

Our results thus provide a proof of concept that the shape of warped throats (as encoded by the warp factor)
may be distinguishable by its effect on the inflationary 
observables.
Interestingly, the same warped geometry can also leave an observable effect on particle physics
observables, e.g., the pattern of soft supersymmetry breaking masses.
Hence, combined data from particle physics and cosmology may allow us to decipher the underlying compactification geometry.
While we explicitly demonstrated these differences by comparing the AdS and KS throats, 
it is straightforward to extend our analysis  to other warped geometries as well.
For example, it would be interesting to study
the signatures of other throats, e.g. the baryonic branch of the KS solution \cite{Baryonic} and other Einstein-Sasaki metrics such as the 
$Y_{p,q}$ \cite{Ypq}, to learn how various warped throats may be 
distinguished from each other.  
Compactifcation effects 
may show up in the CMB for
other stringy scenarios, some issues  
have been explored \cite{Topology}.
Work along these lines is underway.

We thank M.~Huang, I.~Klebanov, L.~McAllister, and H.~Tye for discussions.
We have been informed of a forthcoming paper on related issues \cite{Cornell}.
This work was supported in part by NSF CAREER Award No. PHY-0348093, DOE grant DE-FG-02-95ER40896, a Research Innovation
Award and a Cottrell Scholar Award from Research Corporation.

\end{document}